**Teaching and understanding of quantum interpretations in modern physics courses**

Charles Baily[1] and Noah D. Finkelstein

Department of Physics
University of Colorado
Boulder, CO 80309-0390 USA

**ABSTRACT:** Just as expert physicists vary in their personal stances on interpretation in quantum mechanics, instructors vary on whether and how to teach interpretations of quantum phenomena in introductory modern physics courses. In this paper, we document variations in instructional approaches with respect to interpretation in two similar modern physics courses recently taught at the University of Colorado, and examine associated impacts on student perspectives regarding quantum physics. We find students are more likely to prefer realist interpretations of quantum-mechanical systems when instructors are less explicit in addressing student ontologies. We also observe contextual variations in student beliefs about quantum systems, indicating that instructors who choose to address questions of ontology in quantum mechanics should do so explicitly across a range of topics.

I.  Introduction
II.  Courses Studied
III.  Variations in Instructional Approaches
IV.  Double-Slit Experiment
V.  Variations in Student Perspectives
VI.  Consistency of Student Perspectives
VII.  Discussion and Conclusions

---

[1] Direct correspondence to: Charles.Baily@Colorado.EDU

# I. INTRODUCTION

Introductory courses in classical physics (as well as everyday experience) generally promote in students a *realist* perspective that is both deterministic [1, 2] and local. From this classical point of view, physical quantities such as the position and momentum of a particle have an objective existence independent of experimental observation, and measurements performed on one system cannot affect the outcome of measurements performed on another system that is physically isolated from the first. [3] Such assumptions can be justified when dealing with the macroscopic world, and teaching classical physics in this way has the advantage of appealing to students' everyday intuitions. However, many introductory modern physics students consequently face significant hurdles when they first learn about the decidedly probabilistic and nonlocal theory of quantum mechanics, which precludes any *local realist* interpretation. [4]

In terms of assessing student difficulties in quantum mechanics, several conceptual surveys have been developed, [5-12] though most are appropriate for advanced undergraduate and beginning graduate students since they address topics such as the calculation of expectation values, or the time evolution of quantum states. The *Quantum Physics Conceptual Survey* (QPCS) [12] is the most recently developed assessment instrument designed specifically for introductory modern physics students. The authors of the QPCS found that modern physics students had the most difficulty with six questions which they classified as interpretive; for example, the two survey items with the lowest percentage of correct responses ask whether, "according to the standard (Copenhagen) interpretation of quantum mechanics," light (or an electron) is behaving like a wave or a particle when traveling from a source to a detector. [The authors report that only ~20% of students chose the correct response for each of these two questions.] The QPCS authors also found that not only do a significant number of students perform reasonably well on the non-interpretative questions while still scoring low on the interpretative items, there were no students who scored high on the interpretative questions but scored low on the non-interpretative ones. As the authors note, this parallels a finding by Mazur [13] when comparing student performance on conventional physics problems versus ones requiring conceptual understanding. These results suggest that many introductory modern physics students may grasp how to use the computational tools of quantum mechanics, without a corresponding facility with notions (such as wave-particle duality) that are at odds with their classical intuitions.

Despite its many successes in accounting for experimental observations, quantum theory is still plagued by questions of interpretation more than 80 years after its advent. [14] Among the myriad interpretations of quantum mechanics, the Copenhagen Interpretation is commonly referred to as the accepted, standard interpretation of quantum mechanics; ironically, physicists do not seem to agree on what exactly this "standard" interpretation entails. For the purpose of discussion, Cramer [15] identified five key concepts as a minimal set of principles for the standard Copenhagen Interpretation:

(i) Heisenberg's uncertainty principle (includes the concept of wave-particle duality).
(ii) Born's statistical interpretation (includes the meaning of the state vector given

by the probability law $\mathbf{P = \Psi^*\Psi}$.
(iii) Bohr's concept of complementarity (includes the complementary nature of wave-particle duality; characterizes the uncertainty principle as an intrinsic property of nature rather than a peculiarity of the measurement process).
(iv) Heisenberg's identification of the state vector with "knowledge of the system" (includes the use of this concept to explain the collapse of the state vector).
(v) The positivism of Heisenberg (declining to discuss meaning or reality and focusing interpretive discussions exclusively on observables).

The addition, subtraction or modification of one or more of these principles can lead to other interpretations of quantum mechanics. For example, the Statistical Interpretation [16] criticizes the assumption that a state vector can provide a complete description of individual particles; instead, the state vector encodes probabilities for the outcomes of measurements performed on an ensemble of similarly prepared systems. These interpretations and others (e.g., the Many-Worlds Interpretation, [17] or certain nonlocal hidden-variable theories [18] make identical experimental predictions, yet differ greatly in their ontological implications.

The positivistic aspect of the Copenhagen Interpretation (the refusal to discuss meaning or reality) is arguably one reason why this particular interpretation has maintained such popularity over the years, in that it allows practicing physicists to apply quantum theory without having to worry about what is "really going on" – otherwise known as "Shut Up and Calculate!" [19] Still, a number of experimental tests of the foundations of quantum mechanics in recent decades [20] have inspired some physicists to take what we are calling a *quantum perspective*, by ascribing physical reality to the wave function (in essence, equating the wave function with the system it describes). A recent survey [21] of quantum physics instructors at the University of Colorado and elsewhere (all of whom use quantum mechanics in their research) found that 30% of those surveyed thought of the wave function as a physical matter wave, while nearly half preferred to view the wave function as containing information only; the remaining respondents held some kind of mixed view on the physical interpretation of the wave function, or saw little distinction between the two choices. Only half of those who expressed a clear preference (matter wave or information wave) did so with confidence, and were of the opinion that the other view was probably wrong. In light of this overall state of affairs, it is not surprising that many physicists will choose to take an *agnostic* perspective, recognizing that there are many possibilities without taking a definite stand on which particular interpretation might best correspond with reality.

There are certainly many factors that influence instructors when deciding on what to teach in an introductory modern physics course, and it seems reasonable to assert that the personal beliefs of instructors about the nature of quantum mechanics play an important role in how they choose to address questions of interpretation. In example, we consider comments made in an informal end-of-term interview by a recent modern physics instructor who does not use quantum mechanics in his research as a plasma physicist, and who claimed to personally hold an *agnostic* perspective on quantum physics:

> "It seems like there's a new book about different interpretations of quantum mechanics coming out every other week, so I see this as something that is still up for debate among physicists. When I talked about the double-slit experiment in class, I used it to show students the need to think beyond $F=ma$, but I didn't talk about any of that other stuff. […] We did talk a little about (quantum weirdness) at the very end of the semester, but it was only because we had some time left over and I wanted to give the students something fun to talk about.

Another recent modern physics instructor found that *quantum* interpretations were particularly useful to him in constructing models during his many years of research as an atomic physicist, and was explicit in teaching such a model to his students when discussing the double-slit experiment, by telling them that each electron must pass through both slits simultaneously and interfere with itself on its way from the source to the detector.

When asked at the end of the semester about their preferred interpretation of the double-slit experiment, [1, 2] students from a modern physics course where the instructor was explicit in teaching a *quantum* interpretation overwhelmingly chose to agree with a statement that describes each electron as a delocalized wave packet that propagates through both slits. However, a significant majority of students from a second modern physics course (taught by the agnostic instructor quoted above) said they preferred a *realist* interpretation, agreeing that each electron is a tiny particle that travels through one slit or the other on its way to the detector. Almost every student who had said they preferred a *realist* interpretation of the double-slit experiment also agreed that an electron in an atom must have a definite position at all times, which would again be consistent with a *realist* perspective; nearly half of the students who had preferred the wave-packet description of an electron in the double-slit experiment also agreed with this *realist* statement about atomic electrons. Among the conclusions drawn from these results were: (1) instructors can have a significant impact on student perspectives in contexts where they are explicit about teaching interpretations of quantum processes; and (2) students seem more likely to default to *realist* interpretations of quantum phenomena when instructors are not explicit in promoting an alternative perspective.

The relationship between university instructors' beliefs and their practices in the classroom has been reviewed in detail elsewhere, [22] though Kane *et al*. [23] have been critical of a number of studies that have characterized classroom practices based on the self-reported beliefs and attitudes of instructors, rather than through direct observation. Studies concerning the influence of physics instructors' classroom practices on students have largely focused on the attitudes and beliefs of students about the nature of learning and the nature of science. [24-26] More recently, studies have appeared on how instructional choices impact student perceptions of the nature of classroom activities; [27] and how faculty beliefs may influence which pedagogical tools they employ in their courses, [28] as well as influence how norms are established in the classroom. [29, 30] We are unaware of any prior research specific to the relationship between instructor practices and student beliefs about the nature of quantum mechanics.

Our own prior characterizations of the two instructional approaches discussed above were the result of end-of-term interviews with instructors and a limited number of informal classroom observations; specific course practices relevant to how faculty address questions of interpretation in introductory modern physics courses were not discussed. We therefore document here two modern physics courses recently taught at the University of Colorado, both with similar content and learning environments, but where one instructor was explicit about teaching *quantum* interpretations in the earlier stages of the course, while the instructor for the second course focused primarily on calculation while taking a less explicit and more agnostic approach to questions of interpretation. There are several ways in which we compare the two courses in detail: (1) an analysis of posted lecture slides and classroom observations, in order to document explicit and implicit messages sent to students throughout the semester; (2) interviews with instructors from both courses, to understand why specific instructional choices were made; and (3) pre- and post-instruction survey questions designed to probe student beliefs about quantum mechanics. Through our analysis of these data, we explore the questions of what types of instructional practices might be associated with variations in student perspectives, and whether the perspectives exhibited by students in one context are applied consistently to a context where attention to interpretation was less explicit.

## II. COURSES STUDIED

Each semester, the University of Colorado (CU) offers two versions of an introductory modern physics course (as the third part of a three-semester sequence of introductory courses); one section is intended for engineering majors (PHYS3A) and the other for physics majors (PHYS3B). Historically, the curricula for both versions of the course have been essentially the same, with variations from semester to semester according to instructor preferences, and students are allowed to cross-register (i.e., engineers may receive credit for enrolling in the course for physics majors, and vice-versa). [31] In the fall semester of 2005, a team from the physics education research group at CU introduced a transformed curriculum for PHYS3A that incorporated research-based principles. [32] This included interactive engagement techniques (such as in-class concept questions, peer instruction, and interactive computer simulations [33]), as well as revised content intended to emphasize reasoning development, model building, and connections to real-world problems. These transformations, implemented in PHYS3A during the FA05-SP06 academic year, were continued in FA06-SP07 by another professor from the PER group at CU, who then collaborated in the FA07 semester with a non-PER faculty member to adapt the course materials from PHYS3A into a curriculum for PHYS3B.

The course materials [34] for all five of these semesters (which included lecture slides and concept tests) were made available to the instructors for PHYS3A and PHYS3B from the semester of this study; the instructors for both courses reported changing a majority of the lecture slides to some extent (as well as creating new ones). By examining the course syllabi and categorizing the lecture material for each course into ten standard introductory quantum physics topics, we found the general progression of topics in both classes to be essentially the same (the presentation of content was many times practically identical), with slight differences in emphasis. Table I summarizes the progression of topics from the quantum physics section of the two courses, and the number of lectures spent on each topic. These two modern physics offerings both devoted approximately one-third of the course to special relativity, with the remaining lectures covering the foundations of quantum mechanics and applications to simple systems. Each had a class size of ~75 students, both courses incorporated interactive engagement techniques into lecture, and both used the same textbook [35] from which weekly homework problems were assigned.

| CODE | TOPIC | # OF LECTURES PHYS3A | # OF LECTURES PHYS3B |
|---|---|---|---|
| A | INTRODUCTION TO QUANTUM PHYSICS | 2 | 1 |
| B | PHOTOELECTRIC EFFECT, PHOTONS | 5 | 4 |
| C | ATOMIC SPECTRA, BOHR MODEL | 5 | 3 |
| D | DE BROGLIE WAVES/ATOMIC MODEL | 1 | 1 |
| E | MATTER WAVES/INTERFERENCE | 3 | 2 |
| F | WAVE FUNCTIONS, SCHRODINGER EQUATION | 2 | 5 |
| G | POTENTIAL ENERGY, INFINITE/FINITE SQUARE WELL | 3 | 3 |
| H | TUNNELING, ALPHA DECAY, STM | 2 | 4 |
| I | 3-D SCHRODINGER EQUATION, HYDROGEN ATOM | 4 | 2 |
| J | MULTI-ELECTRON ATOMS, PERIODIC TABLE, SOLIDS | 3 | 3 |

**TABLE I.** Progression of topics and number of lectures devoted to each topic from the quantum physics portion of both modern physics courses.

### III. VARIATIONS IN INSTRUCTIONAL APPROACHES

While the learning environments and progression of topics for both modern physics courses were essentially the same, the two courses differed in sometimes obvious, other times more subtle ways with respect to how each instructor addressed student perspectives and themes of interpretation. An analysis of the instructional materials used in each of the two courses offers a first-pass comparison of the two approaches. The textbook provides some discussion of interpretation (introduced in the context of the double-slit experiment) by addressing the probabilistic interpretation of the wave function, also emphasizing that "which-path" questions in the double-slit experiment are simply unanswerable. Homework is another key avenue by which faculty establish norms regarding which aspects of the course content are deserving of the most attention. When looking at the homework assignments for each course, we found no (or very minimal) opportunities for students to reflect on physical interpretations of quantum phenomena. Similarly, an examination of the midterms and finals from both courses revealed no emphasis on questions of interpretation in quantum mechanics. The one place that afforded the most faculty-student interaction with respect to interpretation was in the lecture portions of each course, and so we examine how faculty specifically addressed questions of interpretation in class.

The first analysis of lecture materials entailed a coding of lecture slides that were used in class and then later posted on the course website. We employed a simple counting scheme by which each slide was assigned a point value of zero or one in each of three categories, according to its relevance to the themes summarized in Table II (which also gives the total count in each category for both modern physics courses). These three categories (denoted as l*ight*, *matter*, and *contrasting perspectives*) were chosen to highlight key lecture slides that were explicit in promoting non-classical perspectives. Since light is classically described as a wave, slides that emphasized its particle nature, or explicitly addressed its dual wave-particle characteristics, were assigned a point in the *light* category; similarly, slides that emphasized the wave nature of matter, or its dual wave-particle characteristics, were given a point in the *matter* category. Other key slides (*contrasting perspectives* category) were those that addressed randomness, indeterminacy, or the probabilistic nature of quantum mechanics; or those that made explicit contrast between quantum results and what would be expected in a classical system. While most of the slides in Table II received only one point in a single category, many slides were relevant to multiple categories, and so the point totals do not represent the total number of relevant slides from each course.

| THEME | DESCRIPTION OF LECTURE SLIDE OR CONCEPT TEST | PHYS3A | PHYS3B |
| --- | --- | --- | --- |
| Light | Relevant to the dual wave-particle nature of light, or emphasizing its particle characteristics. | 15 | 9 |
| Matter | Relevant to the dual wave-particle nature of matter, or emphasizing its wave characteristics. | 15 | 16 |
| Contrasting perspectives | Relevant to randomness, indeterminacy or the probabilistic nature of quantum mechanics; explicit contrast between quantum results and what would be expected classically. | 28 | 22 |

TABLE II. Description of lecture slides relevant to three key themes in promoting non-classical perspectives, and the point totals for each of the two modern physics courses.

PHYS3A had a greater number of slides that scored in the *light* and *contrasting perspectives* categories, though the graphs in Fig. 1 (which group the point totals for each course by topic area, as listed in Table I) show that this difference can be largely attributed to instructor choices at the outset of the quantum physics section of the courses, in topic category B (photoelectric effect and photons). That this topic area should stand out in this analysis seems natural if one considers that: (i) the photoelectric effect demonstrates a need for a particle description of light; (ii) the double-slit experiment with single photons requires both a wave and a particle description of light in order to fully account for experimental observations; and (iii) being the first specific topic beyond the introductory quantum physics lecture(s), it represents an opportunity to frame the content of the course in terms of the need to think beyond classical physics. While both modern physics courses had the greatest point totals in this topic category, PHYS3A devoted a greater portion of lecture time here to addressing themes of indeterminacy and probability (PHYS3A also totaled more points in the *light* category, though this difference can be largely attributed to Instructor A's brief discussion of lasers, a topic not covered in PHYS3B). Figure 2 shows shows the ratio of the point totals for each of the three interpretive themes (from topic area B only) to the total number of slides used during these lectures; the differences between the two courses in terms of the amount of lecture time spent contrasting perspectives is statistically significant ($p = 0.001$, by a one-tailed $t$-test). We note, finally, that in both courses all three of these interpretive themes received considerably less attention at later stages of the course.

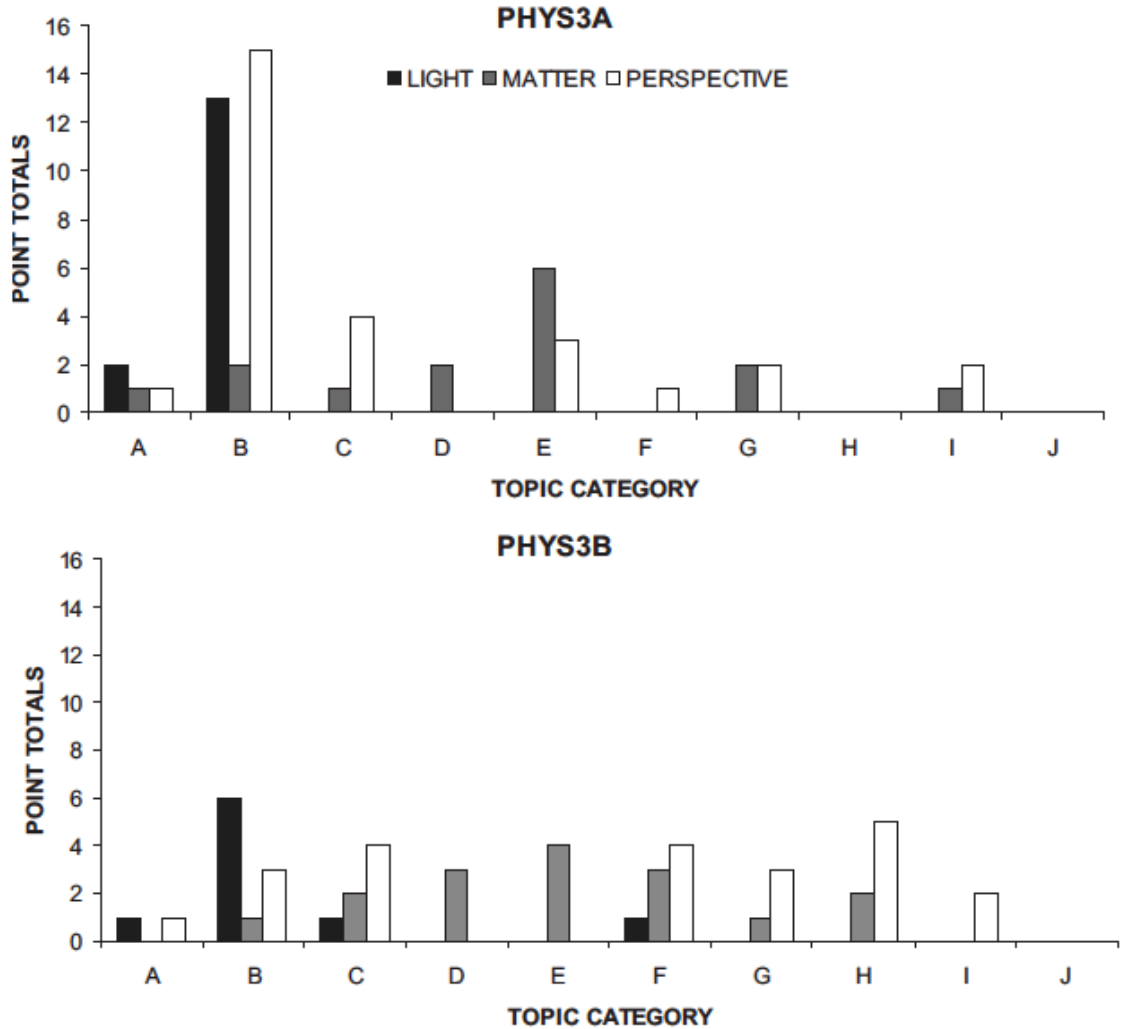

**FIG. 1.** The occurrence of lecture slides for both PHYS3 courses by topic (as listed in Table I) for each of the themes described in Table II.

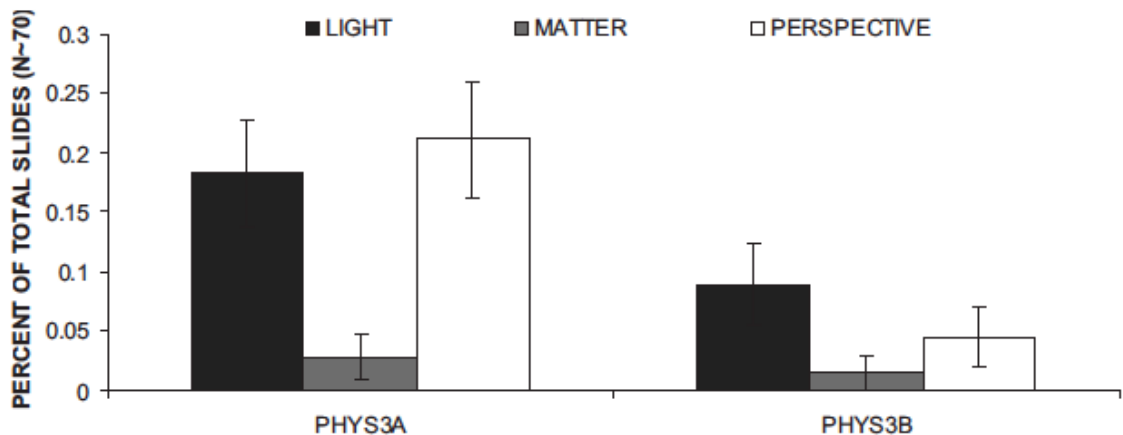

**FIG. 2.** The ratio of point totals from topic area B for each interpretive theme to the total number of slides used during these lectures. Error bars represent the standard error on the proportion.

The lecture slide shown in Fig. 3 is one example of how PHYS3A differed from PHYS3B in attending to student perspectives during the discussion of photons, by explicitly addressing the likelihood for students to think of particles as being localized in space. There were no comparable slides from PHYS3B from this topic category, though this should not be taken to mean that Instructor B failed to address such issues at other times during the semester, or one-on-one with students. We note simply that there were no such explicit messages as part of the artifacts of the course in this topic area (which reflects a value judgment on the part of the instructor regarding content), and PHYS3B students who accessed the lecture slides as posted online would have no indication that such ideas were deserving of any particular emphasis.

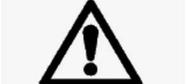

**FIG. 3.** Lecture slide used in PHYS3A during the discussion of photons.

While there are coarse differences in how the instructors addressed student perspectives in some topic areas, the instructional approaches sometimes differed in more subtle ways. The two slides shown in Fig. 4 are illustrative of how the differences between the two courses could sometimes be less obvious, though still of potential significance. Both slides summarize the results for the system referred to in PHYS3A as the *Infinite Square Well*, and by Instructor B as the *Particle in a Box*. At first glance, the two slides may seem almost identical: each depicts the first-excited state wave function of an electron, as well as listing the normalized wave functions and quantized energy levels for this system. Both slides make an explicit contrast between the quantum-mechanical description of this system and what would be expected classically, each pointing out that a classical particle can have any energy, whereas an electron confined in a potential well can only have specific energies. However, PHYS3A differed from PHYS3B by emphasizing a model of the electron as a standing wave, delocalized and spread out, stating explicitly that the electron should not be thought of as bouncing back and forth between the two walls of the potential well. PHYS3B focused instead on the kinetic energy of the system, pointing out that a classical particle can be at rest, whereas the quantum system has a *nonzero* ground-state energy. It is arguable that Instructor B's choice of language, to speak of a "particle in a box" exhibiting zero-point motion, can implicitly reinforce in students the *realist* notion that in this system a localized particle is bouncing back and forth between two potential barriers. Both of these slides received a point in the *contrasting perspectives* category, but only the slide from PHYS3A received a point in the *matter* category for its emphasis on the wavelike properties of an electron in a potential well.

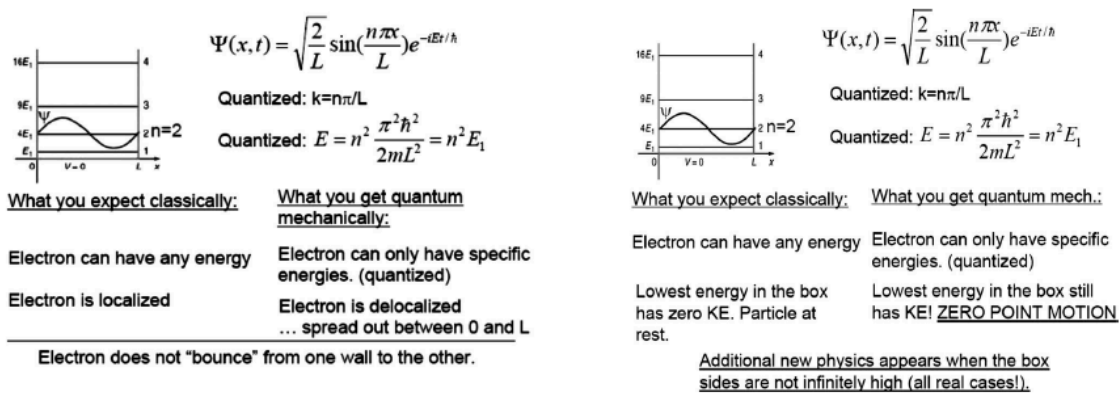

**FIG. 4.** Lecture slides from PHYS3A (left, *Infinite Square Well*) and from PHYS3B (right, *Particle in a Box*).

## IV. DOUBLE-SLIT EXPERIMENT

The double-slit experiment is a natural sub-topic in the discussion of photons, since it requires both a wave and a particle description of quanta in order to completely account for experimental observations. In this experiment, a mono-chromatic beam impinges on two closely spaced slits and diffracts; wavelets spread out behind the slits and interfere in the regions where they overlap, with bright fringes appearing on the detection screen where the antinodal lines intersect. The wave description of quanta explains the interference pattern on the detection screen, while a particle description addresses the fact that individual quanta are detected as localized particles. It was observed that both courses instructed students during lecture on how to relate the distance between the slits and the wavelength of the beam to the locations of the maxima and minima of the interference pattern, and both used the Quantum Wave Interference simulation [36] in class to provide students with a visualization of the process.

Both PHYS3 courses also instructed students that the intensity of the beam can be turned down to the point where only single quanta pass through the apparatus at a time; individual quanta are detected as localized particles on the screen, yet an interference pattern still develops. Observations of several recent offerings of the modern physics courses taught at the University of Colorado have revealed that instructors vary in how to explain this result to students. One interpretation, which was preferred by Instructor A, models individual quanta as delocalized wave packets that propagate through both slits simultaneously, interfere with themselves, and then become localized when interacting with the detector. Instructor A was quite explicit in teaching this model, devoting a great deal of lecture time to a step-by-step explanation of the process.

At one time or other, Instructor B did offer to students the idea of self-interference in the double-slit experiment as one possible interpretation of the observations, but he ultimately emphasized the prevalence of an agnostic stance among practicing physicists. When faced in class with the question of whether an electron has a definite but unknown position, or has no definite position until measured, Instructor B answered:

> "Newton's Laws presume that particles have a well-defined position and momentum at all times. Einstein said that we can't know the position. Bohr said, philosophically, it has no position. *Most physicists today say: We don't go there. I don't care as long as I can calculate what I need* [emphasis added]."

Overall, PHYS3B spent less time addressing questions of interpretation and perspective in comparison to PHYS3A, and students took note of Instructor B's reluctance to address such issues in class, as one student commented after the end of the semester:

> "[This] made me think back to class and asking questions like that, and [Instructor B] kind of blew them off, saying we don't know, it doesn't matter. […] So that's the big picture that comes to mind: Well, we don't really know."

In an end-of-term interview, Instructor B clarified his attitude toward teaching any particular perspective to students in a sophomore-level course:

> "In my opinion, until you have a pretty firm grip on how QM actually works, and how to use the machine to make predictions, so that you can confront the physical measurements with pairs of theories that conflict with each other, there's no basis for [berating] the students about, 'Oh no, the electron, it's all in your head until you measure it.' They don't have the machinery at this point, and so anybody who wants to stand in front of [the class] and pound on the table and say some party line about what's really going on, nevertheless has to recognize that the students have no basis for buying it or not buying it, other than because they're being yelled at."

Instructor A agreed with Instructor B on the role of experiment in assessing a physical theory, but seemed to differ on what he felt students could conclude from these particular experimental observations:

> "This image that [students ]have of this [probability] cloud where the electron is localized, it doesn't work in the double-slit experiment. You wouldn't get diffraction. If you don't take into account [the distance between] both slits and the electron as a delocalized particle, then you will not come up with the right observation, and I think that's what counts. The theory should describe the observation appropriately. […] It really shouldn't be a philosophical question just because there are different ways of describing the same thing [i.e. as a wave or a particle]. They seem to disagree, but in the end they actually come up with the right answer."

## V. VARIATIONS IN STUDENT PERSPECTIVES

In the last week of the semester, students from both PHYS3 courses responded to an online survey designed to probe their beliefs about quantum mechanics. Students received homework credit for responding to the survey (equivalent to the number of points given for a typical homework problem), and the response rate for both courses was approximately 90%. Students were also told they would only receive full credit for providing thoughtful answers, and the text of the survey itself emphasized in bold type that there were no "right" or "wrong" answers to the questions being asked, that we were particularly interested in what the students actually believed. The wording of the items on the survey was vetted ahead of time by instructors for both courses, and interviews conducted after the end of the semester indicated that students interpreted the meaning of the questions in a way that was consistent with the authors' intent.

An essay question from the online survey asked respondents to argue for or against three statements made by fictional students regarding their interpretation of the double-slit experiment with single quanta, as depicted in the *Quantum Wave Interference* simulation (shown in Fig. 5). In this simulation, a bright circular spot representing the probability density for a single electron (A) emerges from a gun, (B) passes through both slits, and (C) a small dot appears on a detection screen; after a long time (many electrons) an interference pattern develops (not shown).

Each of the following statements from the essay question is meant to represent one of three potential perspectives on how to think of the electron between when it is emitted from the gun and when it is detected at the screen; respondents were free to agree or disagree with one, two or all three fictional students.

**Student One** [Realist]: The probability density is so large because we don't know the true position of the electron. Since only a single dot at a time appears on the detecting screen, the electron must have been a tiny particle, traveling somewhere inside that blob, so that the electron went through one slit or the other on its way to the point where it was detected.

**Student Two** [Quantum]: The blob represents the electron itself, since an electron is described by a wave packet that will spread out over time. The electron acts as a wave and will go through both slits and interfere with itself. That's why a distinct interference pattern will show up on the screen after shooting many electrons.

**Student Three** [Agnostic]: Quantum mechanics is only about predicting the outcomes of measurements, so we really can't know anything about what the electron is doing between being emitted from the gun and being detected on the screen.

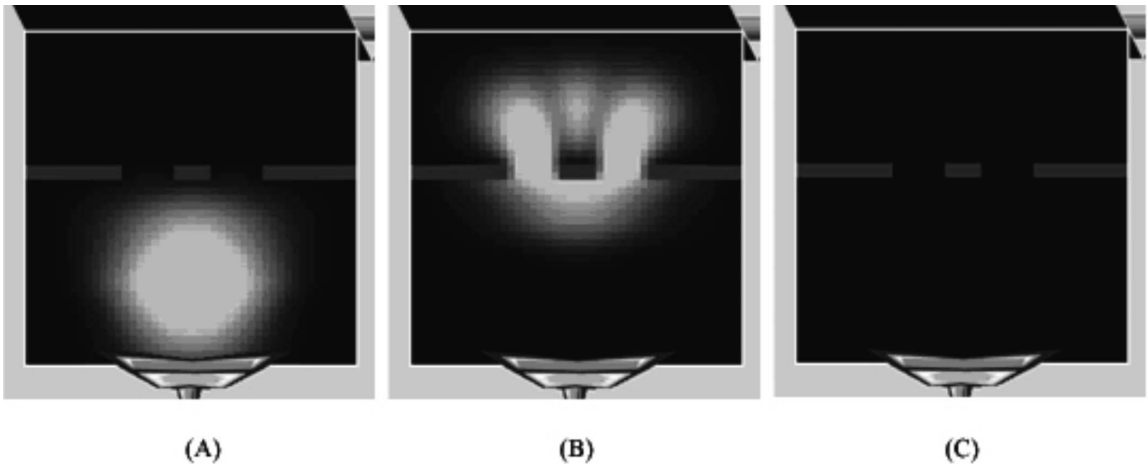

**FIG. 5.** Sequence of screenshots from the Quantum Wave Interference simulation.

The results for both PHYS3A and PHYS3B are shown in Fig. 6, where responses were categorized simply by which fictional student(s) the respondents agreed with (Realist, Quantum, or Agnostic), whatever their reasoning might be. While most students chose to agree with only a single statement, there were a few respondents from both courses who chose to agree with both the fictional Realist and Agnostic students, or with both the Quantum and the Agnostic students; we felt the Realist and Quantum statements were not incompatible with the Agnostic statement, since agreeing with the latter allowed students to acknowledge that they had no way of actually knowing if their preferred interpretations were correct. The relatively few students who responded in this way were grouped together with the other students in the Realist or Quantum categories, as appropriate.

As would be predicted based on the practices of the instructors, most of the students from PHYS3A chose to agree with the Quantum statement (which describes the electron as a delocalized wave packet that interferes with itself), whereas the responses from PHYS3B students were much more varied. Students from PHYS3B were nearly four times more likely to prefer the Realist statement than students from PHYS3A; similarly, PHYS3B students were half as likely to favor the Quantum description. More specifically, 29% of PHYS3B students chose to agree with the Realist statement of Student One, and 27% of PHYS3B students agreed with the Agnostic stance of Student Three, while only a combined 11% of students from PHYS3A chose either of these responses.

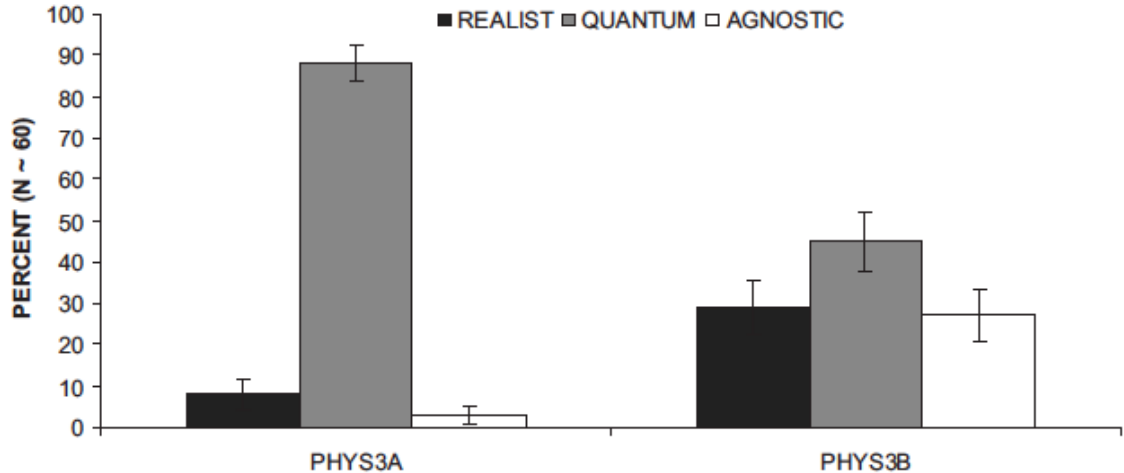

**FIG. 6.** Results for an essay question concerning interpretation in the double-slit experiment for each of the two modern physics courses ($N \sim 60$ for each course; error bars represent the standard error on the proportion). See supplemental material below following the main text for additional data from multiple semesters.

## VI. CONSISTENCY OF STUDENT PERSPECTIVES

As was seen in Fig. 1, both PHYS3 courses paid less explicit attention to student perspectives at later stages of instruction, as when covering the Schrodinger model of hydrogen. In their lecture slides, both courses described an electron in the Schrodinger atomic model as "a cloud of probability surrounding the nucleus whose wave function is the solution of the Schrodinger equation," without further elaboration with respect to interpretation. We were interested in knowing if how students came to think of quanta in the context of the double-slit experiment would be relevant to how they perceived an electron in an atom, particularly when they had not been given the same kind of explicit instruction in this topic area as with the double-slit experiment or the infinite square well in PHYS3A. In addition to the essay question, a pre- and post version of the online survey asked students to respond to the following statement using a five-point scale (ranging from *strong disagreement* to *strong agreement*): "An electron in an atom has a definite but unknown position at each moment of time." Students were also asked to provide the reasoning behind their responses in a textbox following the statement. Disagreement with the statement would be consistent with both a *quantum* or *agnostic* perspective, whereas agreement with the statement would be more consistent with a *realist* perspective. The following student quotes are illustrative of why a student might choose to agree or disagree with the statement:

**AGREE:** "The probability cloud is like a graph method. It tells us where we are most likely to find the electron, but the electron is always a point-particle somewhere in the cloud." [Realist]

**DISAGREE:** "The electron is delocalized until we interfere with the system. It is distributed throughout the region where its wave function is non-zero. An electron only has a definite position when we make a measurement and collapse the wave function." [Quantum]

The pre- and post-instruction responses to this statement for both courses are shown in Fig. 7. Surprisingly, at the end of instruction, students from PHYS3A were just as likely to agree with the statement on atomic electrons as students from PHYS3B, despite the emphasis that PHYS3A gave to thinking of an electron as delocalized in other contexts. Both courses showed a modest (and statistically insignificant) decrease in unfavorable responses to this statement between pre- and post-instruction, yet students from both courses were still more likely to agree than disagree with this statement at the end of the semester.

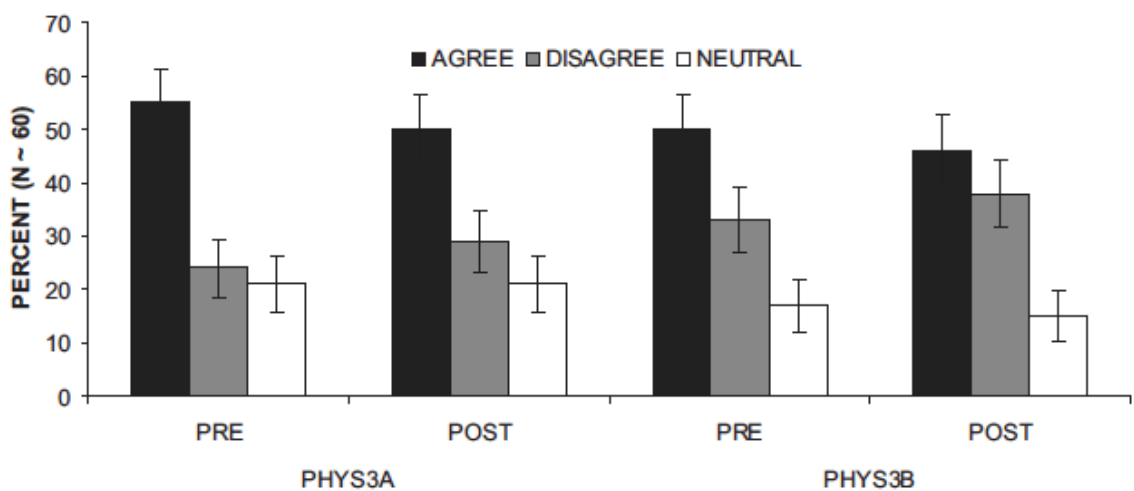

**FIG. 7.** Student responses to an attitudes statement on atomic electrons for each of the two modern physics courses at pre- and post-instruction ($N \sim 60$ for each course and error bars represent the standard error on the proportion). See supplemental material below following the main text for additional data from multiple semesters.

If responses from both courses to the statement on atomic electrons are grouped by how those same students responded to the essay question on the double-slit experiment [Fig. 8] we see that about 70% of the students who preferred a *realist* interpretation in the essay question also chose a response to the statement on atomic electrons that would be consistent with a *realist* perspective. And while students who preferred a *quantum* interpretation in the first essay question were more likely to disagree with the statement on atomic electrons than the students falling into the Realist category, 46% of these students still agreed that an electron in an atom has a definite position at all times. Only in the case of students who preferred the *agnostic* interpretation in the essay question did a majority disagree with this statement, and no students from this group responded neutrally.

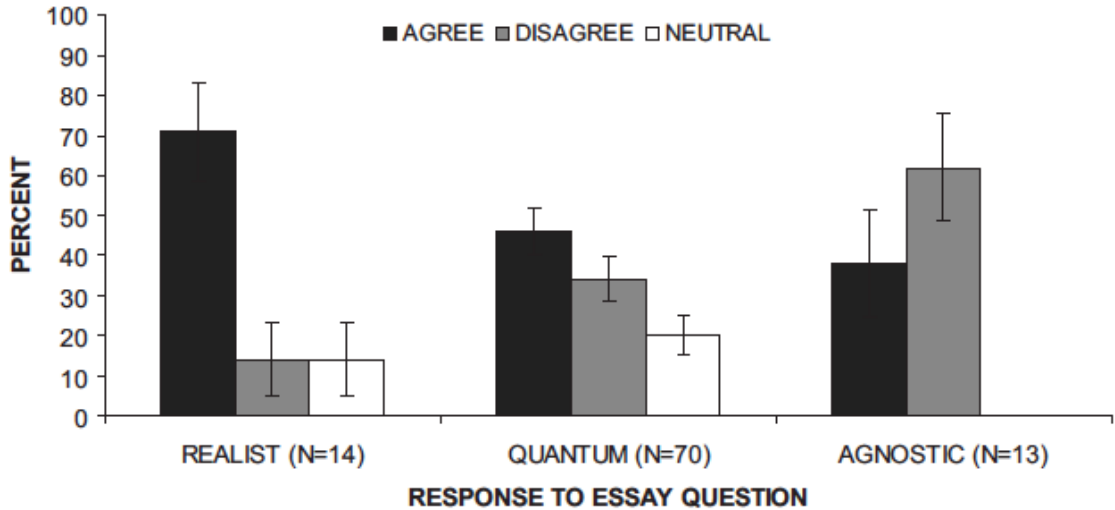

**FIG. 8.** Combined student responses from both courses to the statement on atomic electrons, grouped by how students responded to the essay question on the double-slit experiment (error bars represent the standard error on the proportion).

## VII. DISCUSSION

As discussed in the Introduction, we have observed that modern physics instructors differ not only in their personal perspectives regarding interpretation in quantum mechanics, but also in their decisions to teach (or not teach) about interpretations in their introductory courses. We focus on introductory courses in modern physics because they usually represent a first opportunity for instructors to confront and address students' classical notions of the physical world (even though a majority of students from both PHYS3 courses in this study reported having heard about quantum mechanics in popular venues before enrolling). Later courses that might be relevant to student perspectives (such as upper-division quantum mechanics courses) tend to be even more abstract and mathematically oriented than survey courses, and often leave questions of interpretation completely unaddressed. [20] Here, we have documented two different approaches to teaching interpretation in an introductory modern physics course, and how each approach is associated with significant differences in student responses to survey questions designed to probe students' beliefs about quantum physics.

Our studies and others have indicated that, just as with topics in classical physics, modern physics students are often able to apply mathematical tools without a corresponding conceptual understanding. A major difference between difficulties in classical physics and quantum physics lies in the nature of the questions: when is a particle a particle, and when is it a wave? What is the difference between the experimental uncertainty of classical physics and the fundamental uncertainty of quantum theory? End-of-semester comments from Instructor B support the notion that students who preferred a *realist* interpretation of the double-slit experiment were not doing so from a simple lack of comprehension:

"Some of the students who I considered to be the most engaged went with (the

Realist student): 'The electron is a real thing; it's got to be in there somehow. I know that's not what you told us, but that's what I'm thinking.' I thought that was just great; it was sort of honest. They were willing to recognize that that's not what we're saying, but they're grappling with that's how it's got to be anyways."

Furthermore, one-on-one interviews conducted with students after the end of the semester showed that those who favored a *realist* perspective were still able to correctly describe from memory the double-slit experimental setup and observations. This leads us to conclude that it is unlikely that students who preferred a *realist* interpretation in the double-slit experiment did so because they were unaware of the particulars of the topic.

We also find it worth noting that the instructors in this study, while sometimes explicit in teaching *an* interpretation of quantum processes, were not explicit in teaching these as *interpretations*. In other words, they did not teach quantum mechanics from an axiomatic standpoint, did not explicitly teach the *Copenhagen Interpretation* (or any other formal interpretation), nor did they frame their interpretations of quantum phenomena in terms of *modeling,* or *nature of science* issues. Instead, instructors for both courses addressed questions of interpretation as they arose within the contexts of specific topics, without making, for example, the physical interpretation of the wave function into a major topic unto itself.

When comparing the two courses considered in this study, we see that Instructor A's more explicit approach to teaching one particular interpretation of the double-slit experiment had a significant impact on how students thought of photons and other quanta within that specific context. Instructor B's less explicit and more agnostic instructional approach is reflected in the greater variation of student responses to the essay question, and we note that not only were PHYS3B students more likely than PHYS3A students to prefer an agnostic stance on the double-slit experiment, PHYS3B students were also more likely to align themselves with a *realist* interpretation. Notably, the emphasis given in PHYS3A toward thinking of quanta as delocalized in the absence of measurement in the double-slit experiment and the infinite square well did not seem to transfer to a context where instruction was less explicit in addressing student ontologies. Both courses were similar in their treatment of the Schrödinger atomic model, and student responses from both courses regarding the existence of an electron's position in an atom were not significantly different, with the majority of students from both courses favoring a *realist* perspective in this specific context.

We were able to investigate the consistency of student perspectives across contexts by comparing student responses to the essay question on the double-slit experiment with a statement regarding the position of an electron in an atom. We find that most every student who preferred a *realist* interpretation of the double-slit experiment also took a realist stance on the question of whether an electron in an atom has a definite position. On the other hand, almost half of the students who preferred the wave-packet description of an electron in the double-slit experiment would still agree that an electron in an atom has a definite position at all times. Such responses evidence the greater likelihood for students to hold *realist* perspectives when instruction is less explicit, and suggest that instructors who wish to promote a particular perspective when teaching modern physics should be explicit in doing so across a range of topics, rather than assuming it to be sufficient to address student ontologies primarily at the outset of the course.

We do not advocate any specific approach to teaching interpretation in introductory modern physics courses, but rather note that instructors should be aware of the potential impact they may have on student thinking as a consequence of their instructional choices regarding interpretation. Our studies indicate that students do not come into a course on modern physics as "blank slates" with regard to interpretive ideas, and that instructors who spend less time explicitly attending to student prior knowledge and intuition are less likely to transition students to consistent perspectives that are *not realist*. As our courses currently stand, student perspectives seem to be highly context dependent. Many students have demonstrated mixed perspectives that may seem contradictory to expert physicists, indicating the need for a more detailed exploration of student perspectives in quantum physics beyond the broad characterizations of Realist, Quantum, or Agnostic, which is the subject of current studies.

**ACKNOWLEDGMENTS**

The authors wish to thank the University of Colorado physics faculty members who helped facilitate this research, as well as Chandra Turpen and the other members of the Physics Education Research group at Colorado for their helpful insights. This work was supported in part by NSF CAREER Grant No. 0448176 and the University of Colorado.

Supplementary material for:
*Teaching and understanding of quantum interpretations in modern physics courses*
Charles Baily and Noah D. Finkelstein

The following graphs contain data collected from additional semesters of the PHYS3 courses described in the associated paper. They serve to corroborate the findings of **Figs. 6** and **7**, providing suggestive trends in the relationship between instructional practices and student perspectives on quantum mechanics.

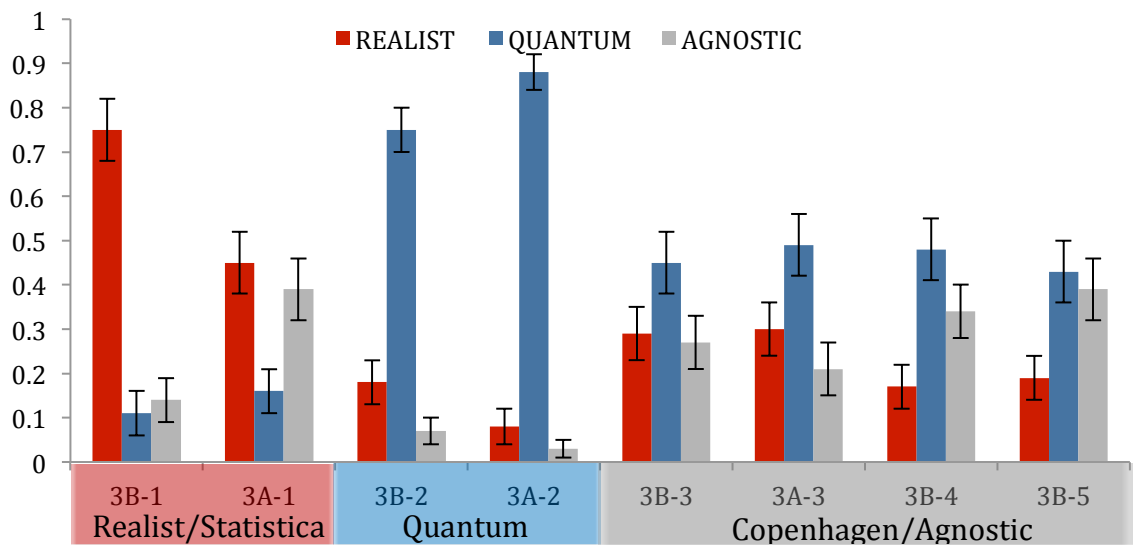

**Figure S1 (color)**: Supplement to **Fig. 6** from main text. This figure shows the distribution of student responses to the double-slit essay question for eight different offerings of the PHYS3A&B modern physics courses; semesters are grouped by color (red, blue, gray) to indicate the instructional approaches for that semester (Realist/Statistical, Quantum, Copenhagen/Agnostic). Instructional approaches were characterized based on classroom observations, faculty interviews, and a review of course materials. Error bars represent the standard error on the proportion. Statistically significant differences in student responses across semesters demonstrate impacts associated with varying instructional approaches. Students from courses taught from a Realist/Statistical perspective were more likely to prefer the Realist interpretation of the double-slit experiment than any of the other modern physics sections. Students from courses where a matter-wave perspective was explicitly taught overwhelmingly chose the Quantum interpretation. Students from the Copenhagen/Agnostic courses are, in general, more evenly split among perspectives, and are among the most likely to prefer the Agnostic perspective.

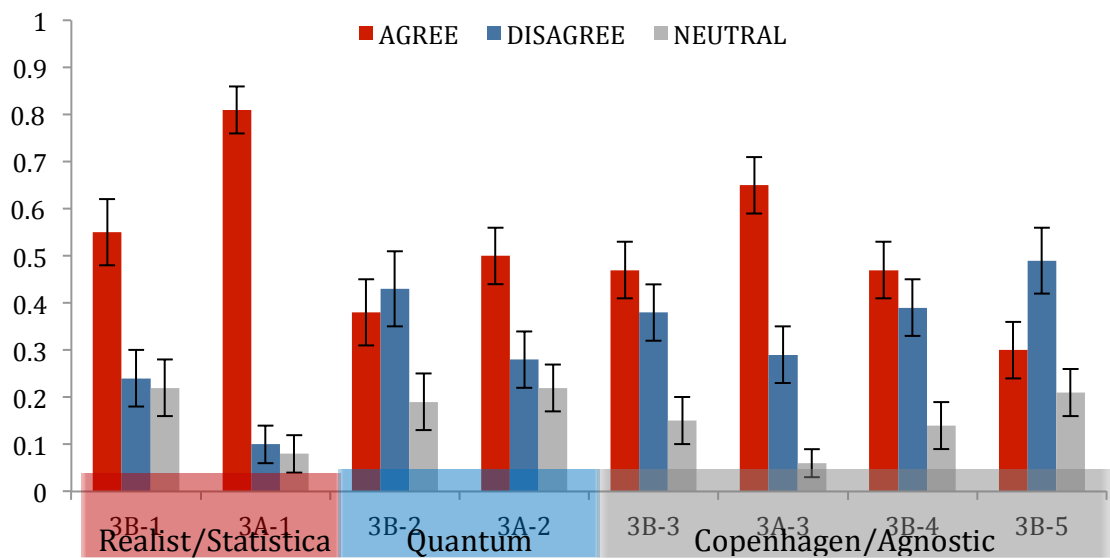

**Figure S2 (color)**: Supplement to **Fig. 7** from main text. This figure shows the distribution of post-instruction student responses from eight different offerings of the PHYS3A&B modern physics to the statement: **"An electron in an atom exists at a definite (but unknown) position at each moment in time."** Semesters are grouped by color (red, blue, gray), as in **Fig. S2**, to indicate the instructional approaches for that semester (Realist/Statistical, Quantum, Copenhagen/Agnostic). Error bars represent the standard error on the proportion. Here, trends associated with instructional approaches are clear for the realist/statistical approaches, but less obvious for the other semesters. Students from the Realist/Statistical courses were more likely to select a response that would be consistent with a Realist perspective on atomic electrons; students from the Quantum and Copenhagen/Agnostic courses were, in general, more evenly split among perspectives.